\documentclass[12pt,hidelinks]{article}
\usepackage{mathrsfs,amsmath,epsfig,amssymb,color,graphicx,bm}
\usepackage{algorithm,algorithmic,natbib,subcaption,hyperref,lineno}
\usepackage[margin=1in]{geometry}
\usepackage[symbol]{footmisc}

\graphicspath{{figures/}}
\allowdisplaybreaks

 \renewcommand{\theequation}{\arabic{section}.\arabic{equation}}
 \renewcommand{\thefigure}{\arabic{section}.\arabic{figure}}
\newtheorem{theorem}{Theorem}
\newtheorem{lemma}{Lemma}
\newtheorem{remark}{Remark}
\newtheorem{assumption}{Assumption}

\newcommand{\bSigma}{\boldsymbol \Omega}
\newcommand{\bbeta}{\boldsymbol \beta}
\newcommand{\fbeta}{\breve{\boldsymbol \beta}_{\mbox{\tiny \rm  DSE}}}
\newcommand{\hbeta}{{\boldsymbol \beta_0}}
\newcommand{\tbeta}{\breve{\boldsymbol \beta}}

\newcommand{\blue}{\color{black}}
\newcommand{\bPsi}{\boldsymbol \Psi}
\newcommand{\bUpsilon}{\boldsymbol \Upsilon}

\newcommand{\X}{\mathbf{X}}
\newcommand{\sumn}{\sum_{i=1}^{n_k}}

\newcommand{\appsi}{\pi_{ik}^{\text{}*}}
\newcommand{\appsj}{\pi_{jk}^{*}}

\begin{document}
\centerline {\Large\bf DsubCox:  A Fast Subsampling Algorithm for Cox Model }
\centerline {\Large\bf    with Distributed and Massive Survival Data}
\vspace*{0.2in}
\centerline{ {  Haixiang Zhang$^{1*}$, Yang Li$^{2}$ and  HaiYing Wang$^{3}$}}
 \vspace*{0.2in}

\centerline{\small \it $^{1}$Center for Applied Mathematics and KL-AAGDM, Tianjin University, Tianjin 300072, China}
\centerline{\small \it $^{2}$Department of Biostatistics and Health Data Science, Indiana University School of Medicine}
\centerline{\small \it and Richard M. Fairbanks School of Public Health, Indianapolis, IN 46202, USA}
\centerline{\small \it $^{3}$Department of Statistics, University of Connecticut, Storrs,
Mansfield, CT 06269, USA}

\footnotetext[1] {Corresponding author. Email: haixiang.zhang@tju.edu.cn}
\vspace{1cm}

\begin{abstract}
To ensure privacy protection and alleviate computational burden, we propose a
fast subsmaling procedure for the Cox model
with massive survival datasets from multi-centered, decentralized sources.  The
proposed estimator is computed based on optimal subsampling probabilities that
we derived and enables transmission of subsample-based summary level statistics
between different storage sites with only one round of communication. For inference, the asymptotic properties
of the proposed estimator were rigorously established.  An extensive simulation
study demonstrated that the proposed approach is effective. The methodology was
applied to analyze a large dataset from the U.S. airlines. Finally, we offer a user-friendly R function {\tt DsubCox} to facilitate the implementation of our algorithm in practical applications.

  
\end{abstract}

\section{Introduction}

The ubiquity of vast data volumes and privacy preserving needs are becoming
defining characteristics when analyzing massive survival data from
multi-centered, decentralized sources. With limited computational resources,
traditional statistical tools using centralized data are no longer feasible to
address either challenge. For effective processing with big data, subsampling is
a popular statistical technique that can significantly alleviate both
computational and storage burdens.  In recent years, subsampling-based methods for analyzing survival data with large-scale datasets have been extensively explored in the literature.  For example, \cite{zuo2021-sim} studied the sampling-based methods for analyzing massive survival data using an additive hazards model. \cite{SIM-AFT-sampling} and \cite{AFT-2024} considered the optimal
subsampling algorithms for accelerate failure time
models. \cite{rare-Cox-2023} and \cite{JCGS-Cox} proposed optimal subsampling
procedures for Cox regression in the context of rare or non-rate
events. Comprehensive reviews on general subsampling-based big data analysis can
be found in \cite{OSP-review-2021} and \cite{SP-review}, which also collects
subsampling-based regression with non-survival data (\citeauthor{Wang2018-JASA},
\citeyear{Wang2018-JASA}; \citeauthor{Wang2019-JASA}, \citeyear{Wang2019-JASA};
\citeauthor{Wangtzhang-2022}, \citeyear{Wangtzhang-2022};
\citeauthor{HTYZ-2020-AOS}, \citeyear{HTYZ-2020-AOS}; \citeauthor{Wang-MA2020},
\citeyear{Wang-MA2020}).  For privacy preservation with decentralized data,
distributed learning has been gaining increasing popularity as a solution that
utilizes subsample-based summary level statistics only. Several recent works
indicate that distributed learning in combination with subsampling techniques
can balance the needs for privacy protection and maximized utilization of
informative samples (\citeauthor{Zhang-Wang2021}, \citeyear{Zhang-Wang2021};
\citeauthor{Zuo-2021-CS}, \citeyear{Zuo-2021-CS};
\citeauthor{Poisson-JASA-2021}, \citeyear{Poisson-JASA-2021}).  However, there
is a paucity of research on analysis of large-scale for survival data that
leverages the benefits of both subsampling and distributed learning.

In this paper, we propose an approach for Cox regression that integrates
subsampling and distributed learning techniques.  The main idea involves
conducting optimal subsampling on individual datasets separately, and
subsequently, constructing a distributed subsample estimator using summary-level
statistics from multiple datasets. 
Our approach can effectively address the issue of ``isolated data island" in big
data by accommodating multiple datasets, rather than being restricted to single
datasets alone. In addition, the proposed estimator enables privacy-preserving
transmission of subsample-based summary level statistics between different
storage sites with only one round of communication.

The remainder of this paper is organized as follows: In Section \ref{sec2}, we
introduce notation and definitions associated with the Cox model. In Section
\ref{sec3}, we  propose a distributed subsample estimator with
rigorously derived asymptotic properties. In Section \ref{sec5}, we demonstrate
the effectiveness of our approach through simulation studies and an illustrative
example for real-world applications. Some concluding remarks are provided in
Section \ref{sec6}, and detailed proofs are presented in the Appendix.

\section{Model and Notation}\label{sec2}
 \setcounter{equation}{0}

Let $T$ denote the failure time, $C$ represent the censoring time, and $\X$ be a
$p$-dimensional vector of covariates. Given $\X$, we assume that $T$ and $C$ are
conditionally independent. The observed failure time is $Y = \min(T, C)$ with
the failure indicator $\Delta = I(T \leq C)$, where $I(\cdot)$ represents the
indicator function. The Cox's proportional hazard regression model
\cite[]{Cox1972} assumes that the conditional hazard rate function of $T$ given
$\X$ is
\begin{align}\label{01}
  \lambda(t|\X)= \lambda_0(t) \exp(\bbeta^\prime \X),
\end{align}
where $\lambda_0(t)$ represents an unknown baseline hazard function,
$\bbeta = (\beta_1,\cdots,\beta_p)^\prime$ is a $p$-dimensional vector of
regression parameters with its true value is constrained within a compact set
$\Theta \subset \mathbb{R}^p$.

Assume that we observe $K$ large survival datasets denoted as
$\mathcal{D}^{[k]}= \{(\X_{ik}, \Delta_{ik}, Y_{ik}), i=1,\cdots,n_k\}$, where
$n_k$ represents the sample size of $\mathcal{D}^{[k]}$ for $k=1,\cdots,K$, and
the variable $K$ is a bounded variable (\citeauthor{Larry-FACE-2022},
\citeyear{Larry-FACE-2022}; \citeauthor{SIM-FL-2023},
\citeyear{SIM-FL-2023}). The sample size $n_k$ is considerably large, which may impose a substantial computational burden on each local site. {\blue In addition, to license the combination of data across $K$ sources, we require that all observed variables in $\mathcal{D}^{[k]}$'s adhere to an identical distribution. i.e.,  the $K$ distributed local datasets are homogeneous, which can be characterized by the same model (\ref{01}).}

 The presence of such large-scale and distributed survival datasets poses two primary challenges that need to be addressed. The first challenge is how to alleviate the computational burden resulting from a massive sample size, while the second challenge pertains to sharing information across different data sources with privacy protection.  Our idea involves deriving a local estimator from a sample subset at each data source, followed by aggregating these local estimators into a final weighted average. This approach only requires one round of communication with each dataset center, ensuring privacy protection.

\section{Distributed Subsample Estimation}\label{sec3}
\subsection{Subsampling Algorithm}
First, we assign sampling probabilities $\{\pi_{ik}\}_{i=1}^{n_k}$ to the $k$th
dataset $\mathcal{D}^{[k]}$ with $\sum_{i=1}^{n_k} \pi_{ik} = 1$ and $\pi_{ik} > 0$,
where  $k=1,\cdots,K$.  Draw a random subsample of size $r_k$ with replacement based on
$\{\pi_{ik}\}_{i=1}^{n_k}$ from the $k$th
dataset $\mathcal{D}^{[k]}$, where $r_k$ is much smaller than $n_k$. 
Let
$\mathcal{D}^{*[k]} = \{ (\X_{ik}^*, \Delta_{ik}^*, Y_{ik}^*,
\pi_{ik}^*)\}_{i=1}^{r_k}$ denote the selected subsample with size $r_k$
from dataset $\mathcal{D}^{[k]}$, where $\X_{ik}^*$ represents
the covariate, $\Delta_{ik}^*$ the failure indicator, $Y_{ik}^*$ the observed
failure times, and $\pi_{ik}^*$ the inclusion probability. Following \cite{JCGS-Cox} , we construct a
weighted subsample score
function as follows:
\begin{align}\label{SS24}
  \dot{\ell}_k^*(\bbeta) &=  -\frac{1}{n_k} \sum_{i=1}^{r_k} \frac{1}{\pi_{ik}^*} \int_0^\tau \{ \X_{ik}^* - \bar{\X}_k^{*}(t, \bbeta)\}dN_{ik}^*(t),
\end{align}
where
$N_{ik}^*(t) = I(\Delta_{ik}^* =1, Y^*_{ik} \leq t)$,
\begin{eqnarray}\label{X-bar-S}
  \bar{\X}_k^{*}(t, \bbeta) = \frac{S_k^{*(1)}(t, \bbeta)}{S_k^{*(0)}(t, \bbeta)}
\end{eqnarray}
and
\begin{align*} {S}_k^{*(v)}(t,\bbeta) =& \frac{1}{n_k} \sum_{i=1}^{r_k}
  \frac{1}{\pi_{ik}^*}I(Y_{ik}^{*} \geq t) (\X_{ik}^{*})^{\otimes v} \exp
  (\bbeta^\prime \X_{ik}^{*}),~~v=0,1~{\rm and}~2.
\end{align*}

In practical implementation, it is necessary to specify the probabilities $\pi_{ik}$'s when conducting subsampling. Here, we adopt the optimal subsampling probabilities as stated in Algorithm 1 of \cite{JCGS-Cox}, which are also denoted as $\pi_{ik}$'s for convenience. 
The subsample-based estimator ${\breve{\bbeta}_k}$ can be obtained by solving the equation $ \dot{\ell}_k^*(\bbeta) = 0$, where $\dot{\ell}_k^*(\bbeta)$ is defined in (\ref{SS24}) for $k=1,\cdots,K$. 

 The asymptotic normality of ${\breve{\bbeta}_k}$ is presented in
  Lemma~\ref{Lemma1} of the Appendix.  We can estimate the variance-covariance matrix of $\breve{\bbeta}_k$ by
\begin{align}\label{SE1}
  \breve{\mathbf{\Omega}}_k  =\breve{\mathbf{\Psi}}_k^{-1}\breve{\mathbf{\Gamma}}_k\breve{\mathbf{\Psi}}_k^{-1}
\end{align}
 using the subsample $\mathcal{D}^{*[k]}$, where
\begin{align}\label{Hessian-21}
  \breve{\mathbf{\Psi}}_k
  &= \frac{1}{n_k}\sum_{i=1}^{r_k}  \frac{\Delta_{ik}^*}{\appsi} \left[\frac{S_k^{*(2)}(Y_{ik}^*,\breve{\bbeta}_k)}{S_k^{*(0)}(Y_{ik}^*, \breve{\bbeta}_k)} -  \left\{\frac{S_k^{*(1)}(Y_{ik}^*, \breve{\bbeta}_k)}{S_k^{*(0)}(Y_{ik}^*, \breve{\bbeta}_k)}\right\}^{\otimes 2}  \right],
  \end{align}
  and
  \begin{eqnarray}
  \breve{\mathbf{\Gamma}}_k
  &=&\frac{1}{n_k^{2}}\sum_{i=1}^{r_k}\frac{1}{\{\appsi\}^2} \left[\int_0^\tau  \{\X^*_{ik} - \bar{\X}_k^{0*}(t, \breve{\bbeta}_k)\}d \hat{M}_{ik}^*(t,\breve{\bbeta}_k)\right]^{\otimes 2}\\
  &&- \frac{1}{n_k^{2}}\sum_{i=1}^{r_k}\frac{1}{\appsi} \left[\int_0^\tau  \{\X^*_{ik} - \bar{\X}_k^{0*}(t, \breve{\bbeta})\}d \hat{M}_{ik}^*(t,\breve{\bbeta})\right]^{\otimes 2},\nonumber
  \end{eqnarray}
together with $S_k^{*(v)} (Y_i^*,\breve{\bbeta}) = {n_k}^{-1} \sum_{j=1}^{r_k} {\appsj}^{-1}I(Y_{jk}^{*} \geq Y_{ik}^{*})
\X_{jk}^{*\otimes v}\exp(\breve{\bbeta}^\prime \X^{*}_{jk})$ for
$v=0,1,2$; The term $\bar{\X}_k^{0*}(t, {\breve{\bbeta}_k})$ is defined in a similar manner as Equation (16) in \cite{JCGS-Cox}, and
$d\hat{M}^*_{ik}(t,\breve{\bbeta}_k) = dN_{ik}^*(t) - I(Y^*_{ik} \geq
t)\exp(\breve{\bbeta}_k^\prime \X^*_{ik})d\hat{\Lambda}_{0k}^{\mbox{\tiny\rm
    UNIF}}(t,\breve{\bbeta}_k)$. The expression of $\hat{\Lambda}_{0k}^{\mbox{\tiny\rm    UNIF}}(t,\breve{\bbeta}_k)$ follows a similar definition as Equation (17) in the paper by \cite{JCGS-Cox}.

\begin{figure}[htp]
  \centering
  \begin{subfigure}{0.9\textwidth}
    \includegraphics[width=\textwidth]{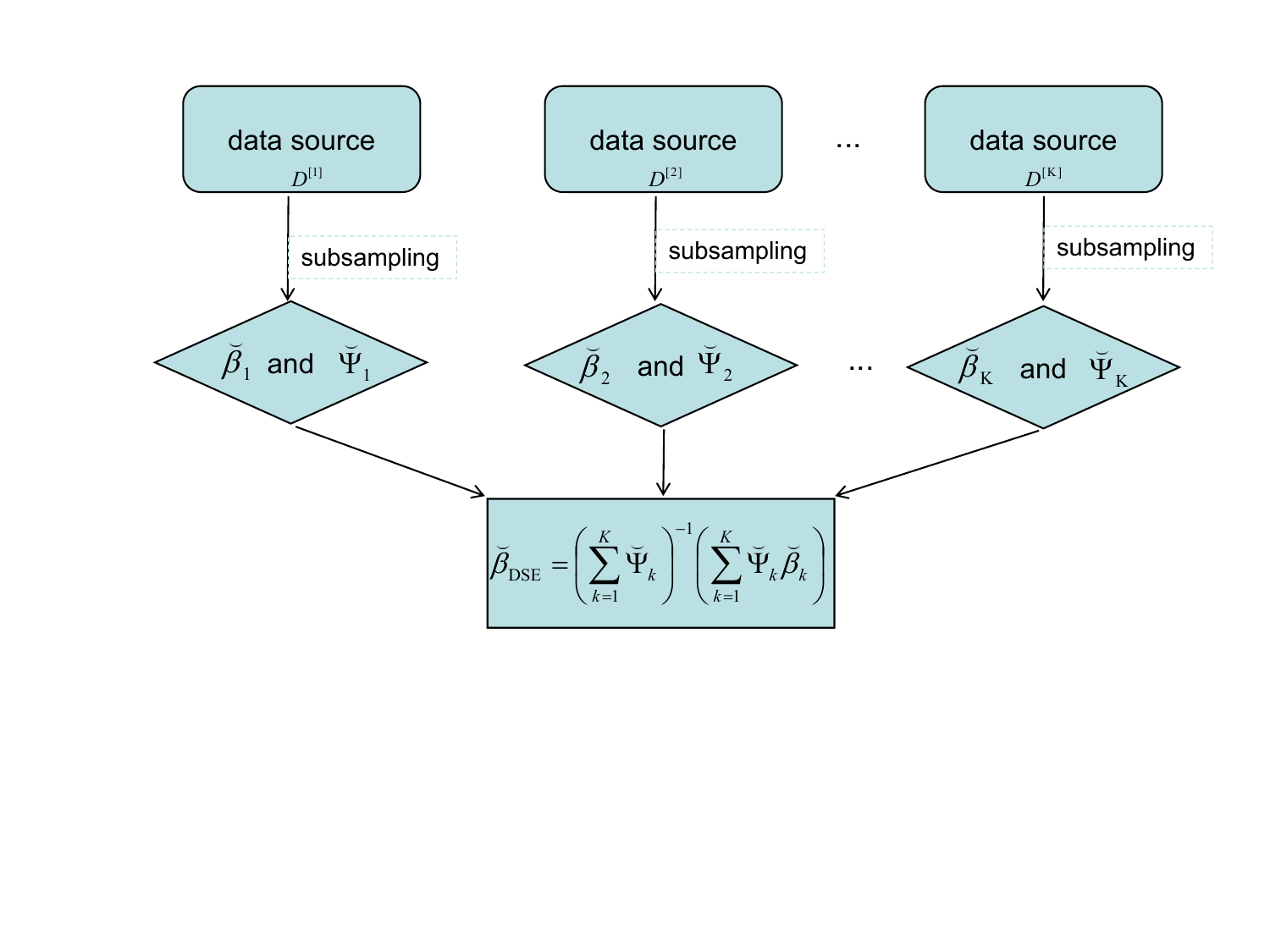}
  \end{subfigure}
 \vspace{-3.9cm}
\begin{center}
\caption{ A diagram of the distributed subsampling algorithm for Cox model.}\label{fig:DSE}
\end{center}
\end{figure}

{ Based on the subsample estimator
  $\breve{\bbeta}_k$'s}, we further introduce a distributed subsample estimator
for privacy-preserving purposes (Figure \ref{fig:DSE}), represented by:

\begin{align}\label{fed-est}
  \fbeta = \left(\sum_{k=1}^K \breve{\mathbf{\Psi}}_k\right)^{-1}\left(\sum_{k=1}^K \breve{\mathbf{\Psi}}_k\breve{\bbeta}_k\right),
\end{align}
 where $\breve{\bbeta}_k$
represents the two-step subsample estimator, and the Hessian matrix
$\breve{\mathbf{\Psi}}_k $ is defined in (\ref{Hessian-21}).  For clarity, we
present the procedure of obtaining $\fbeta$ in Algorithm \ref{algo2} below.


\begin{algorithm}[H]
  \caption {Distributed Subsampling Estimation}\label{algo2}
  $\bullet$ Assign expected subsample sizes $r_k$'s to each of the $K$
  distribued sites.  Run Algorithm 1 of \cite{JCGS-Cox} separately on each site to obtain
  the subsample estimator $\breve{\bbeta}_k$, where $k=1,\cdots,K$.

  $\bullet$ { The subsample estimators $\breve{\bbeta}_k$'s, along with
    $\breve{\mathbf{\Psi}}_k$'s and $\breve{\mathbf{\Gamma}}_k$'s should be
    transmitted to a central site. The distributed subsample estimator $\fbeta$
    is constructed according to the formula presented in (\ref{fed-est}).}

  $\bullet$ Obtain the asymptotic distribution of $\fbeta$, and conduct
  statistical inference tasks. e.g., hypothesis testing and confidence
  intervals.
\end{algorithm}

\subsection{Asymptotic Properties}

To ensure privacy-preserving, the calculation of $\fbeta$ requires summary-level
statistics only instead of individual-level data. Since the estimation procedure
requires only one round of communication between $K$ datasets, it ensures
efficient communication in practical applications and resolves the issue
of ``isolated data island" in big data analysis. The asymptotic property of
$\fbeta$ is established in the following theorem.

\begin{theorem}\label{Th4}
  Under Assumptions~\ref{assu1}-\ref{assu3}, $r_k=o(n_k)$  and with bounded $K$, as
  $n_k\rightarrow \infty$ and $r_k\rightarrow \infty$, we
  have
  \begin{align}\label{Eq4}
    \breve{\boldsymbol \Omega}_{\mbox{\tiny \rm  DSE}}^{-1/2}(\fbeta - \bbeta_0) \stackrel{d}{\longrightarrow} N(\mathbf{0},\mathbf{I}),
  \end{align}
  where $\stackrel{d}{\longrightarrow}$ denotes convergence in distribution, 
  $\breve{\boldsymbol \Omega}_{\mbox{\tiny \rm DSE}} = (\sum_{k=1}^K
  \breve{\mathbf{\Psi}}_k)^{-1} (\sum_{k=1}^K
  \breve{\mathbf{\Gamma}}_k)(\sum_{k=1}^K \breve{\mathbf{\Psi}}_k)^{-1}$, and
  $\mathcal{D}_n = \cup_{k=1}^K \mathcal{D}^{[k]}$.
\end{theorem}

{\blue 
 \begin{remark}
Note that variability in the full sample estimate is negligible compared to the variability
in the subsample since the full sample is so large. Therefore, the asymptotic normality of $\fbeta$ is valuable in conducting statistical inference for the true parameter of the Cox model.
 \end{remark}

\begin{theorem}\label{Theorem2}
 Under Assumptions~\ref{assu1}-\ref{assu3},  if the number of data sources $K$ is bounded,  then we have
 \begin{eqnarray*}
\|\fbeta - \bbeta_0\|\leq K \|\breve{\bbeta}_{k_0} - \bbeta_0\|,
 \end{eqnarray*}
 where $k_0 = \arg\max_{1\leq k \leq K} \{\|\breve{\bbeta}_{k} - \bbeta_0\|\}$.
\end{theorem}
\begin{remark}
To quantify the efficiency loss (the gap between the distributed subsample estimator and the individual data pooling subsample estimator),  we denote $\tilde{\bbeta}$ as the subsampling-based estimator \cite[]{JCGS-Cox} with subsample size $\sum_{k=1}^K r_k$ from the pooling full data $\mathcal{D}_n = \cup_{k=1}^K \mathcal{D}^{[k]}$. From \cite{JCGS-Cox}, the distance between $\tilde{\bbeta}$  and $\bbeta_0$ satisfies $\|\tilde{\bbeta} - \bbeta_0\|= O_{P}(\{\sum_{k=1}^K r_k\}^{-1/2})$. Therefore, we have 
\begin{eqnarray*}
\|\fbeta - \tilde{\bbeta}\| &\leq & \|\fbeta - \bbeta_0\| + \|\tilde{\bbeta} - \bbeta_0\|\\
&\leq & K\cdot O_{P}(r_{k_0}^{-1/2}) + O_{P}\Big(\Big\{\sum_{k=1}^K r_k\Big\}^{-1/2}\Big).
\end{eqnarray*}
That is to say, the efficiency loss between $\fbeta$ and $\tilde{\bbeta}$ becomes negligible as the subsample sizes $r_k$ tend to infinity.
\end{remark}
}

\section{Numerical Studies}\label{sec5}
 \setcounter{equation}{0}
 \setcounter{figure}{0}
\subsection{Simulation}
A simulation study was conducted to validate the estimation and computation
efficiency of the proposed distributed subsampling estimator. We
generated the failure times $T_{ik}$'s from Cox's model with a baseline hazard
function $\lambda_0(t) = 0.5 t$ and the true parameter
$\bbeta_0 = (-1,-0.5,0,0.5,1)^{\prime}$ with 500 replications, dimension $p=5$,
$K=4$, $n_k = 10^6$, and considered four cases for generating the
covariate $\X_{ik}$:\\
{\it Case} \uppercase \expandafter {\romannumeral 1} : $\X_{ik}$ followed a multivariate normal distribution: $\X_{ik} \sim N(\mathbf{0},\bUpsilon)$, where $\Upsilon_{js}=0.3$ if $j\neq s$, and $\Upsilon_{js}=1$ if $j=s$.\\
{\it Case} \uppercase \expandafter {\romannumeral 2}: $\X_{ik}$ followed a mixed multivariate normal distribution: $\X_{ik}\sim 0.5N(-\mathbf{1},\bUpsilon)+0.5N(\mathbf{1},\bUpsilon)$, where $\Upsilon_{js}=0.5^{|j-s|}$.\\
{\it Case} \uppercase \expandafter {\romannumeral 3}: each component of $\X_{ik}$ followed an independent exponential random variables with the probability density function $f(x)=2e^{-2x}I(x>0)$.\\
{\it Case} \uppercase \expandafter {\romannumeral 4}: $\X_{ik}$ followed a
multivariate $t$ distribution with degree of freedom 10, mean zero and
covariance matrix $\bUpsilon$ where $\Upsilon_{js}=0.5^{|j-s|}$.

\begin{table}[htp] 
  \begin{center}
    \caption{Simulation results of the subsample estimator $\breve{\beta}_1$ with CR = $20\%$.}
    \label{tab:1}
    \vspace{0.1in} \small
    \begin{tabular}{lccccccccccc}
      \hline
      & &  & \multicolumn{4}{c}{OSP} &  & \multicolumn{4}{c}{UNIF} \\
      \cline{4-7}\cline{9-12}
      & $r_k$ & &Bias &ESE &SE &CP & & Bias &ESE &SE &CP\\
      \hline
      Case I
      & 200& & 0.0059 & 0.0623 & 0.0629 & 0.954 && -0.0051 &0.0729 & 0.0783 & 0.970 \\
      & 400& & 0.0022 & 0.0425 & 0.0438 & 0.944 &&-0.0029 & 0.0534 & 0.0556 & 0.952 \\
      & 600& & 0.0026 & 0.0354 & 0.0354 & 0.946 &&-0.0031 &0.0429 &  0.0453 & 0.966 \\
      & 800& & 0.0025 & 0.0317 & 0.0306 & 0.956 &&-0.0017 & 0.0388 & 0.0393 & 0.962 \\
      \hline
      Case II
      & 200& & 0.0049 & 0.0582 & 0.0607 & 0.964 && 0.0036 & 0.0768 & 0.0765 & 0.934 \\
      & 400& & 0.0025 & 0.0403 & 0.0422 & 0.960 &&-0.0007 & 0.0520 & 0.0537 & 0.958  \\
      & 600& & 0.0037 & 0.0333 & 0.0343 & 0.958 &&-0.0023 & 0.0422 & 0.0439 & 0.960 \\
      & 800& & 0.0035 &0.0294  &0.0295  & 0.950 && 0.0005 & 0.0368 & 0.0378 & 0.948 \\
      \hline
      Case III
      & 200& & 0.0069 & 0.0765 & 0.0781 & 0.962 && 0.0082 & 0.1032 & 0.1108 & 0.950 \\
      & 400& & 0.0068 &0.0535 & 0.0536 & 0.934 &&  0.0014 &  0.0675 &  0.0748 & 0.968 \\
      & 600& & 0.0051 & 0.0415 & 0.0436 &0.964 && -0.0004 & 0.0548 &  0.0627 & 0.968  \\
      & 800& & 0.0033 & 0.0362 & 0.0378 & 0.960 && 0.0007 &0.0483 & 0.0533 & 0.954 \\
      \hline
      Case IV
      & 200& & 0.0096 & 0.0444 & 0.0454 & 0.954 && -0.0033 &0.0548 & 0.0589 & 0.958 \\
      & 400& &0.0055 & 0.0308 & 0.0313 & 0.952 && -0.0025 & 0.0372 & 0.0410 & 0.970 \\
      & 600& &0.0031 &  0.0245 & 0.0253 &0.960 &&  -0.0024 & 0.0304 & 0.0329 & 0.960 \\
      & 800& &0.0027 &0.0215 & 0.0218 & 0.968 && 0.0002 &  0.0257 &  0.0285 & 0.964 \\
      \hline
    \end{tabular}
  \end{center}
\end{table}
Additionally, we generated the censoring times $C_i$'s from a uniform
distribution over $(0,c_0)$, where $c_0$ is selected to achieve censoring rates
(CR) of approximately $20\%$ and $60\%$, respectively. The pilot sample sizes
were $r_{0k}=200$, and the subsample sizes were chosen to be
$r_k= 200, 400, 600$ and 800, where $k=1,\cdots,5$.

\begin{figure}[htp]
  \centering
  \begin{subfigure}{0.45\textwidth}
    \includegraphics[width=\textwidth]{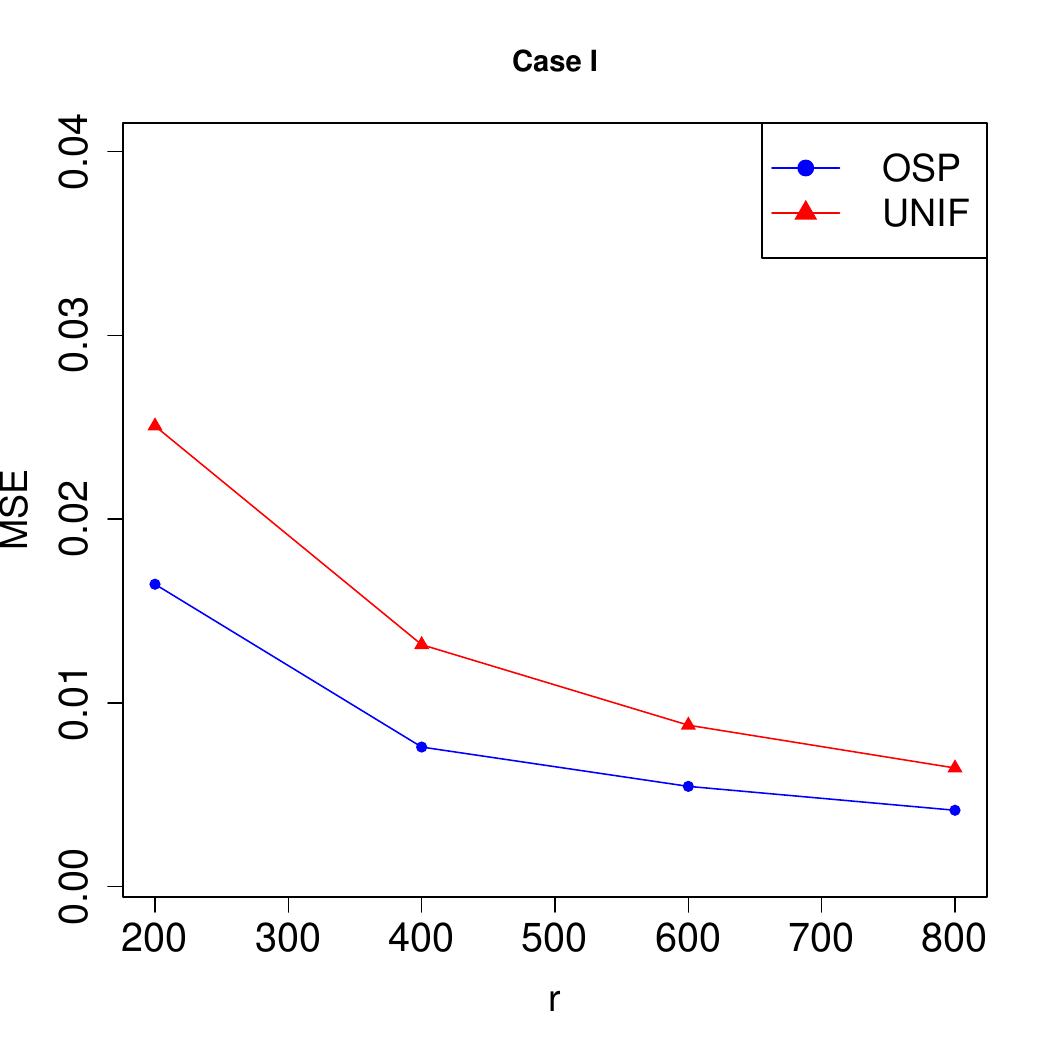}
  \end{subfigure}
  \begin{subfigure}{0.45\textwidth}
    \includegraphics[width=\textwidth]{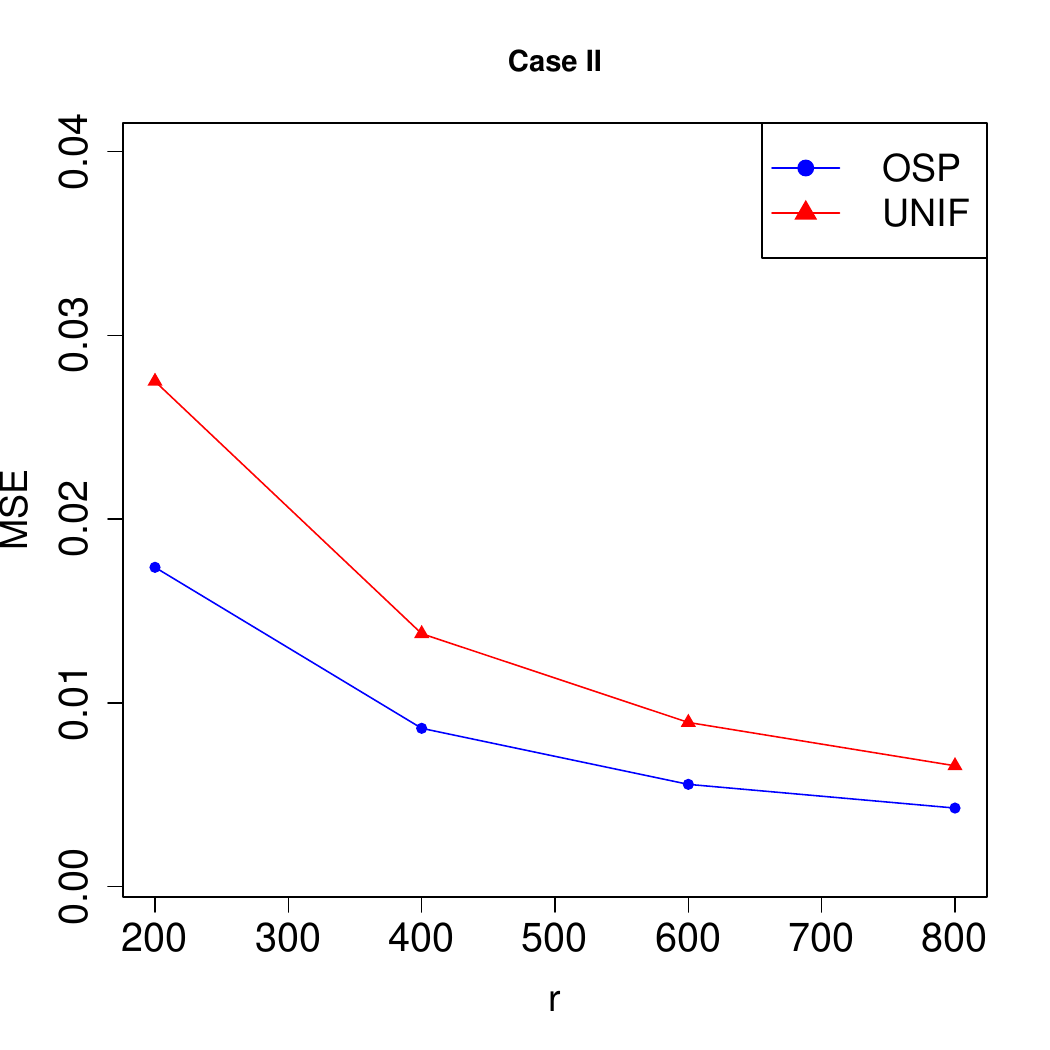}
  \end{subfigure}
    \begin{subfigure}{0.45\textwidth}
    \includegraphics[width=\textwidth]{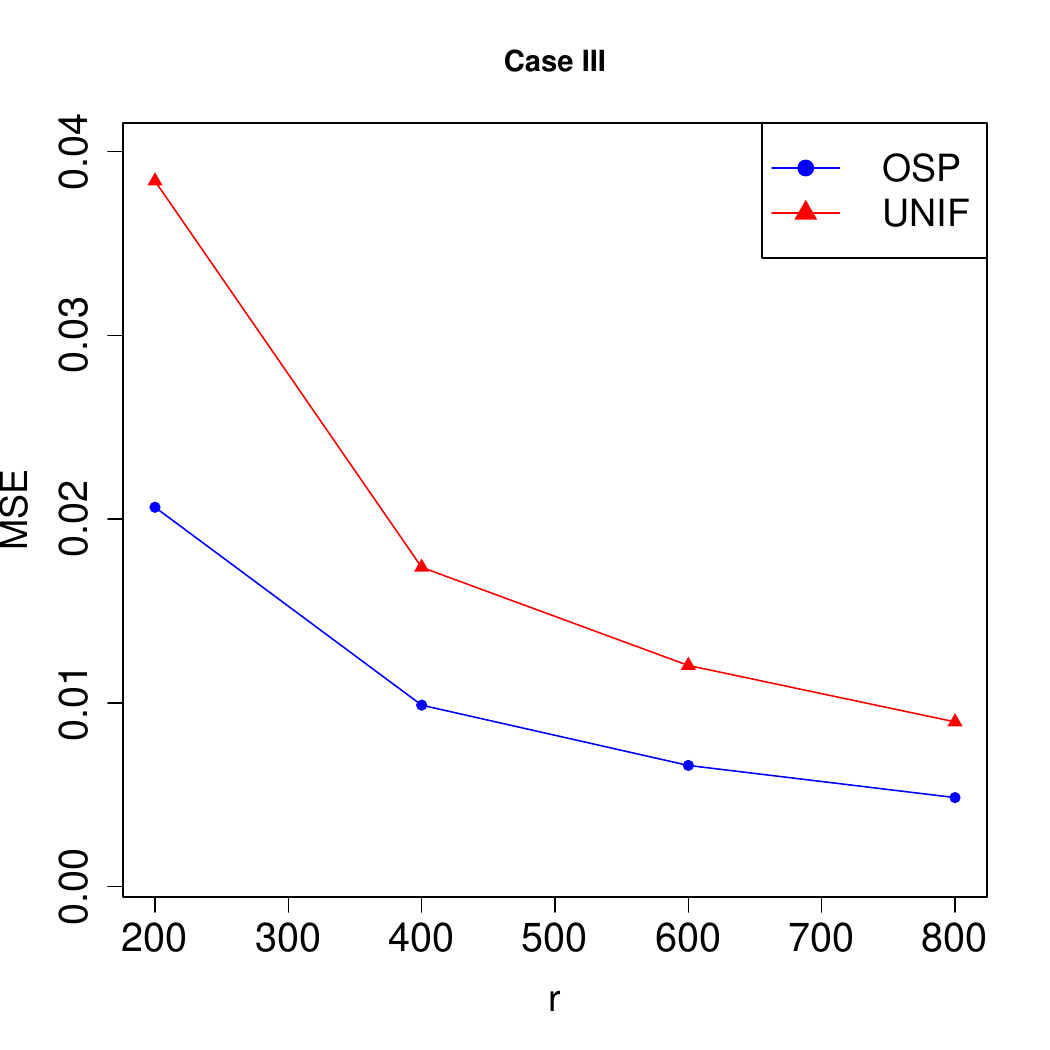}
  \end{subfigure}
  \begin{subfigure}{0.45\textwidth}
    \includegraphics[width=\textwidth]{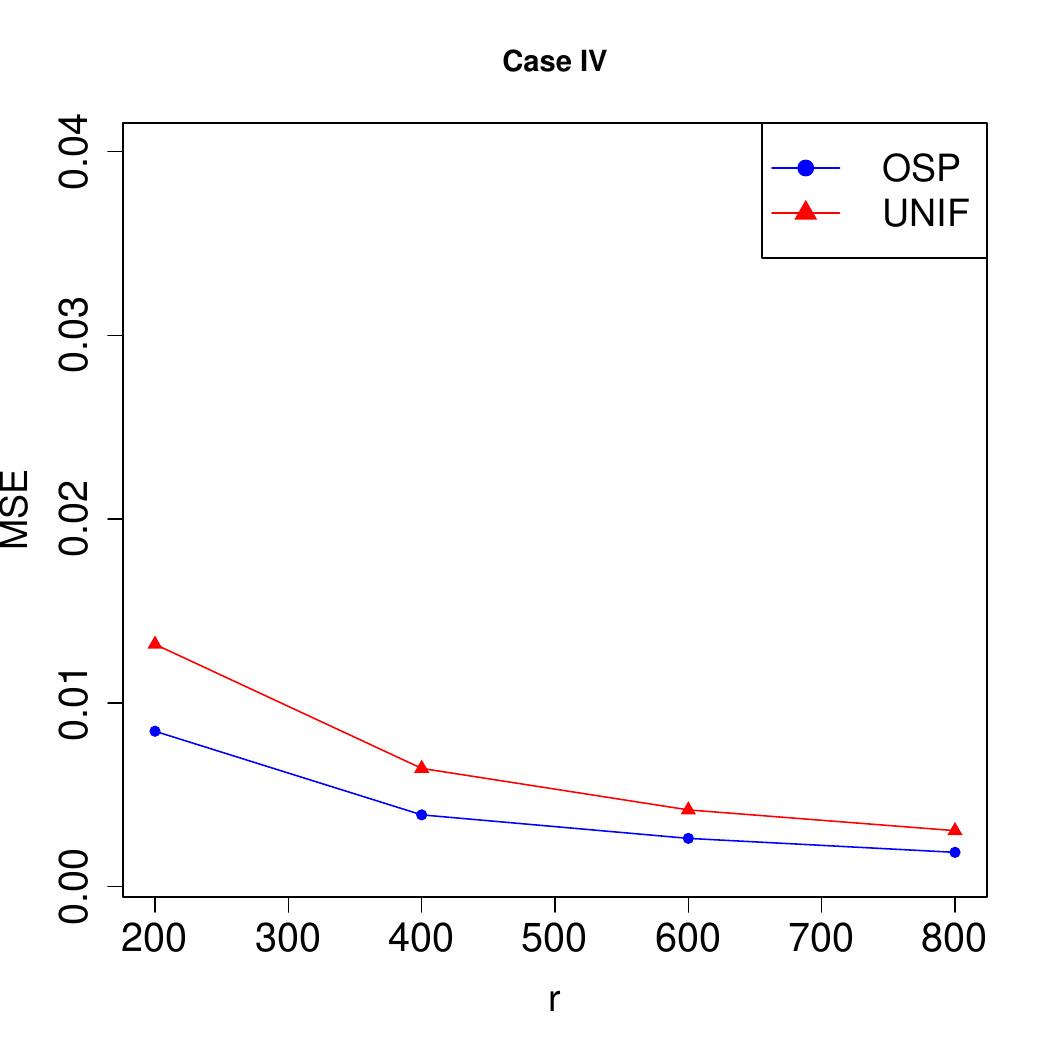}
  \end{subfigure}
 \vspace{-0.1cm}
\begin{center}
\caption{ The MSEs of UNIF and OSP estimators with CR=20\%.}\label{fig:1}
\end{center}
\end{figure}

\begin{figure}[htp]
  \centering
  \begin{subfigure}{0.45\textwidth}
    \includegraphics[width=\textwidth]{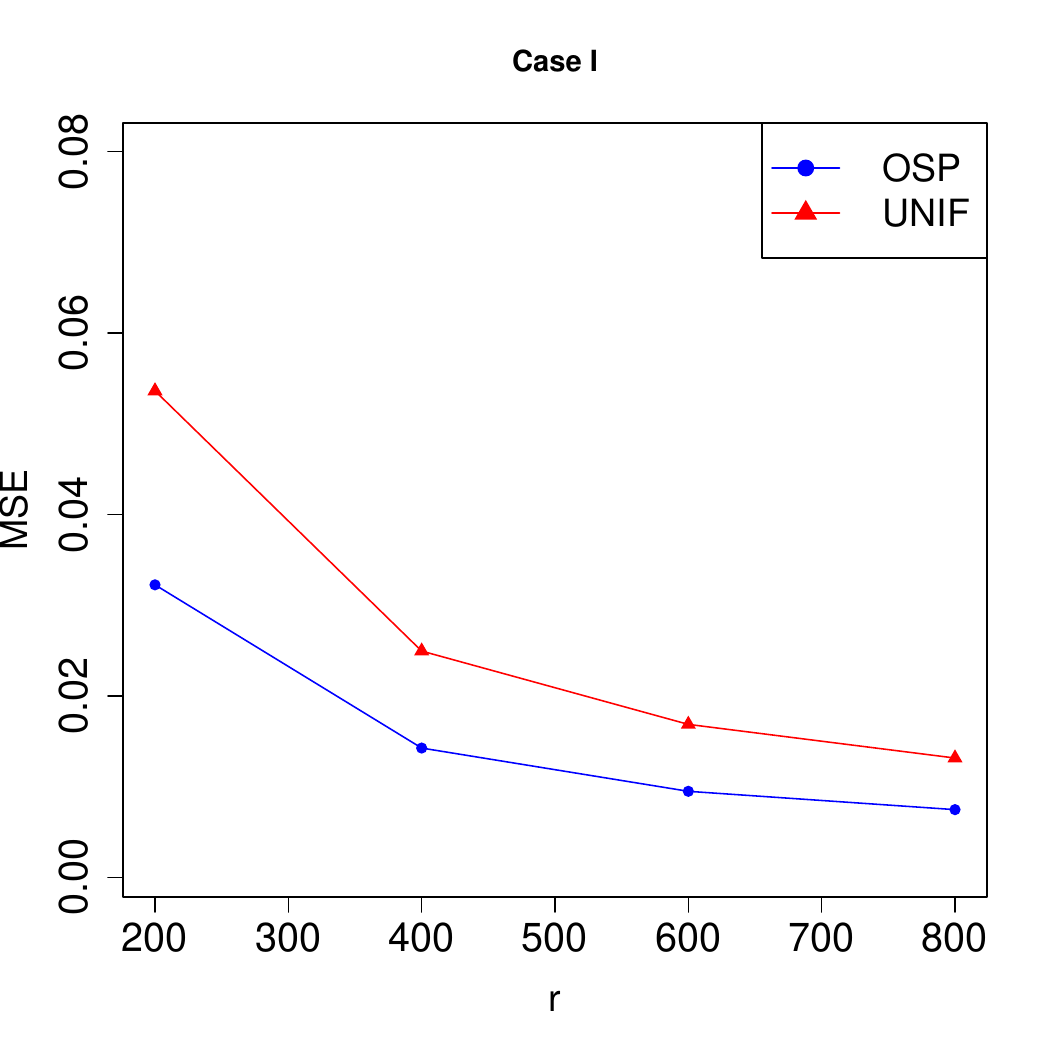}
  \end{subfigure}
  \begin{subfigure}{0.45\textwidth}
    \includegraphics[width=\textwidth]{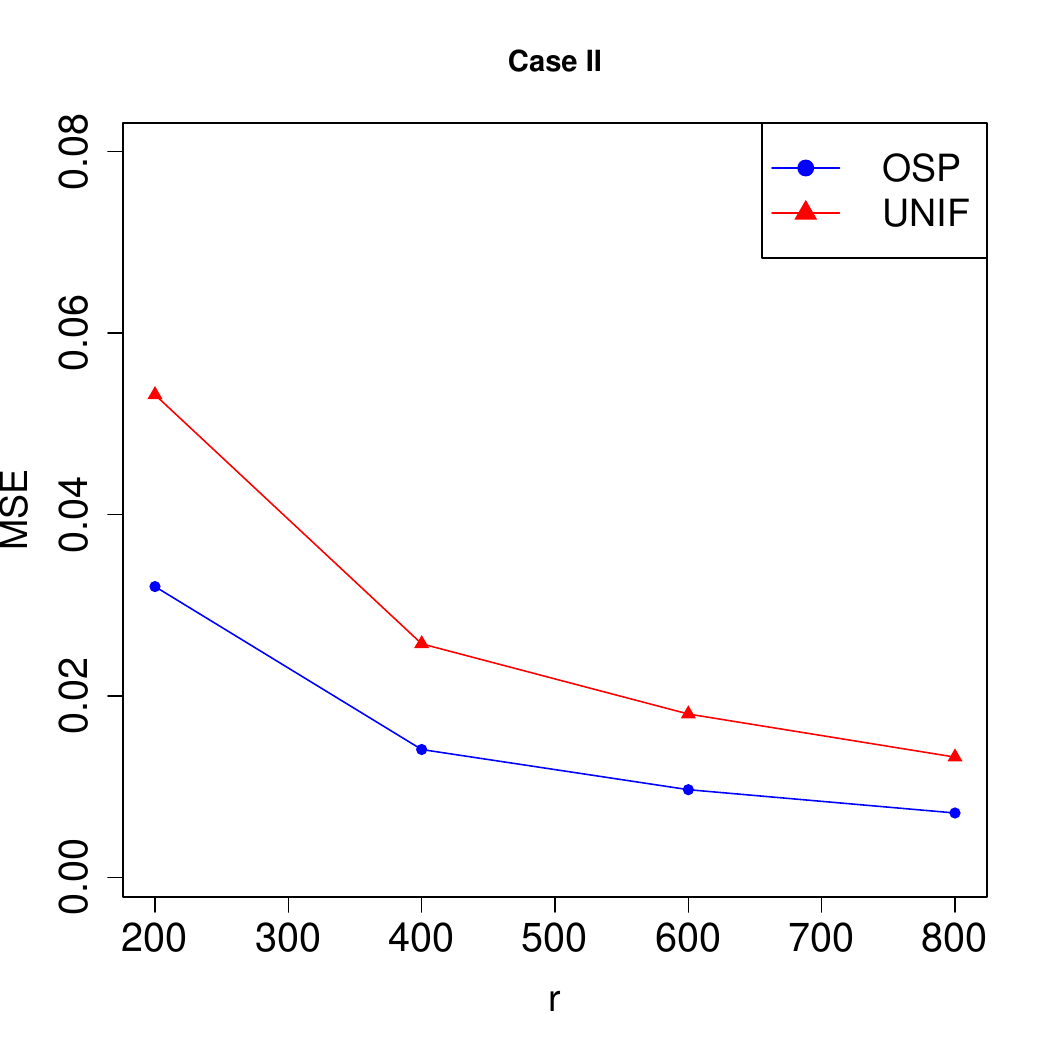}
  \end{subfigure}
    \begin{subfigure}{0.45\textwidth}
    \includegraphics[width=\textwidth]{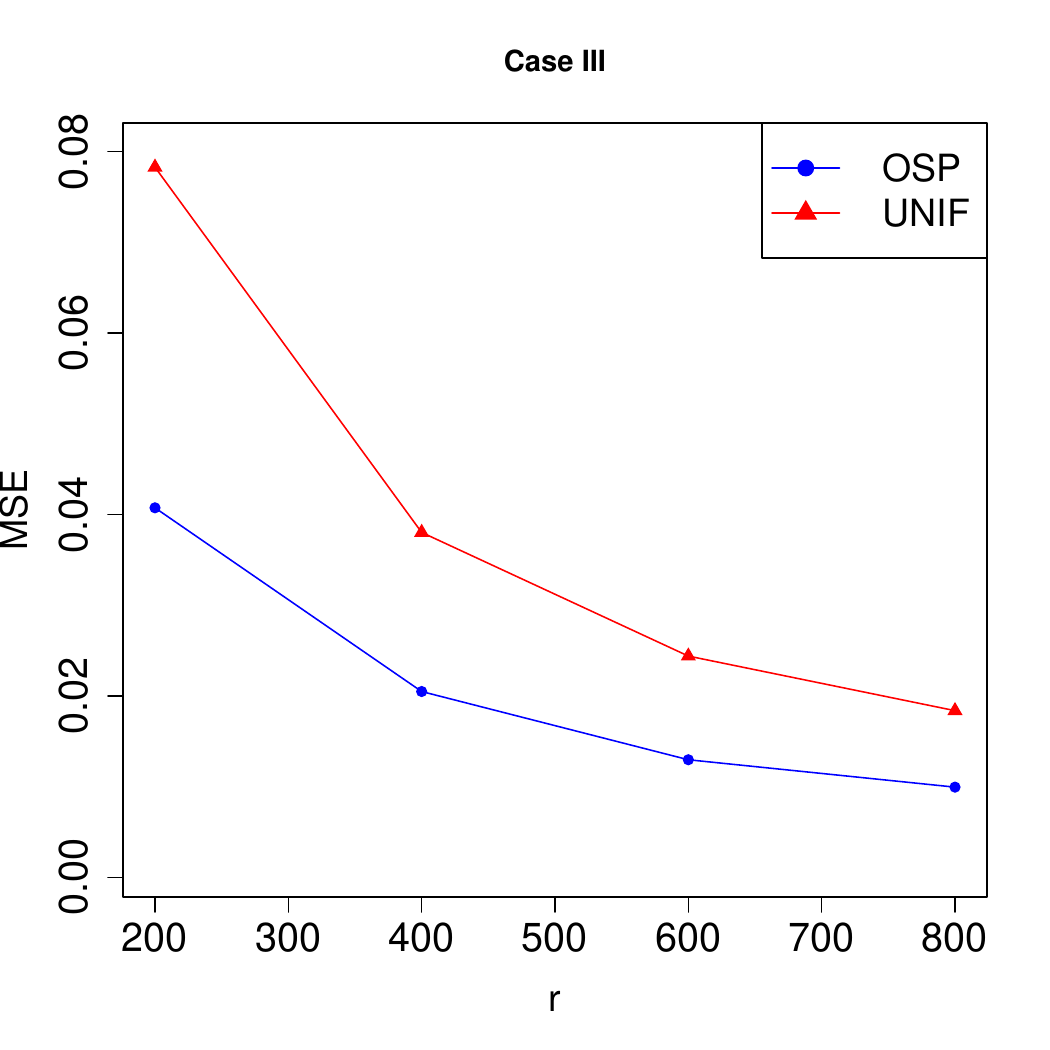}
  \end{subfigure}
  \begin{subfigure}{0.45\textwidth}
    \includegraphics[width=\textwidth]{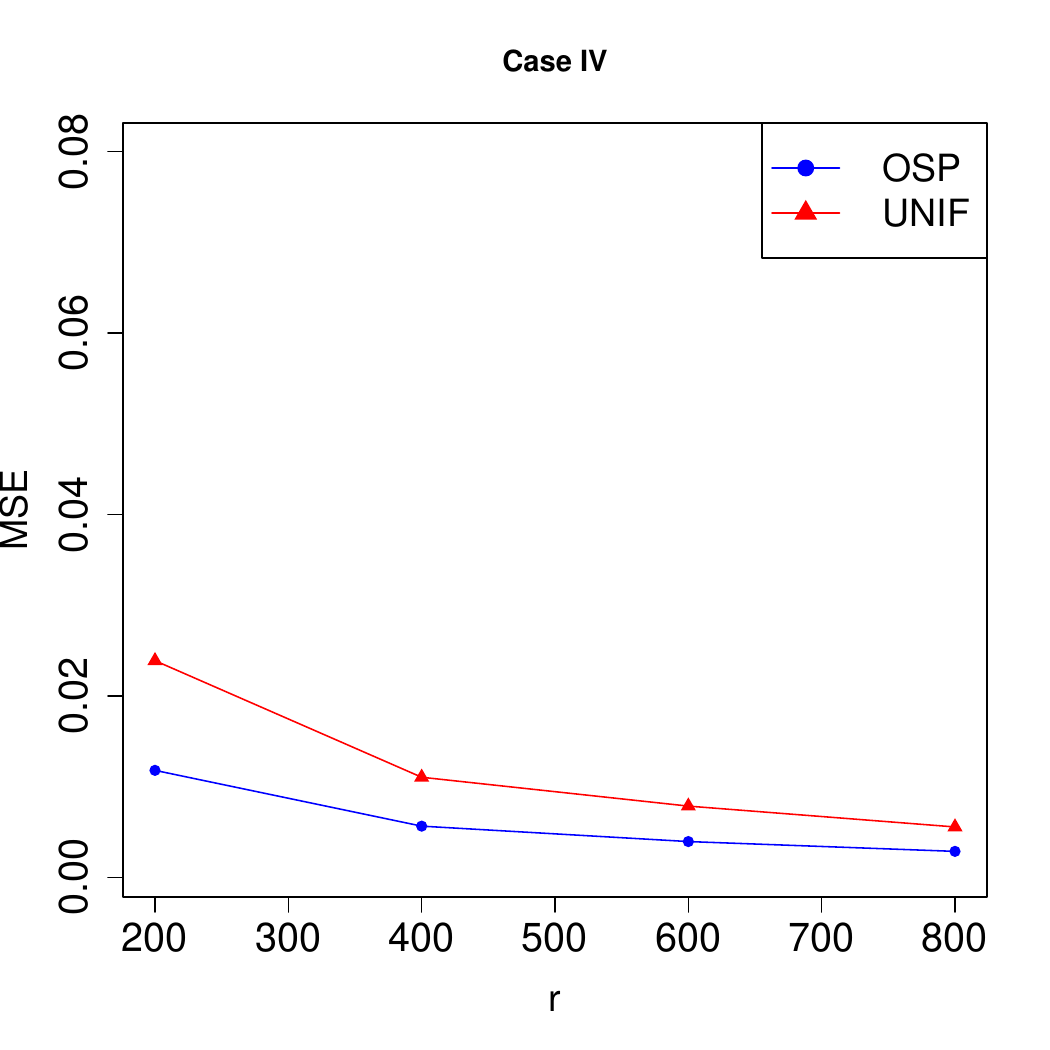}
  \end{subfigure}
 \vspace{-0.1cm}
\begin{center}
\caption{ The MSEs of UNIF and OSP estimators with CR=60\%.}\label{fig:2}
\end{center}
\end{figure}

We compared the distributed optimal subsampling estimator from
Algorithm~\ref{algo2} (``OSP estimator'') with the estimator from the
distributed uniform subsampling method (``UNIF estimator'') and set
$\delta=0.1$. The estimation results for $\beta_1$ were presented in Tables
\ref{tab:1} and \ref{tab:2}, including the empirical biases (Bias), the mean
estimated standard errors (SE), the empirical standard errors (ESE), and the
empirical 95\% coverage probability (CP). The estimation results for other
regression coefficients were similar and were therefore omitted. One can tell
both the OSP and UNIF yielded unbiased estimation. The SE and ESE were close to
each other, and the coverage probabilities agreed well with the asymptotic
normal level. As the sampling size increased, both the SE and ESE became
smaller. {\blue With the same subsample sizes, however, the OSP yielded smaller ESE as compared to the UNIF.} In addition, we calculated the empirical mean
squared error (MSE), defined by
\begin{align*} {\rm MSE}(\fbeta) = \frac{1}{500}\sum_{b=1}^{500} \|\fbeta^{(b)}
  - \bbeta_0\|^2,
\end{align*}
where $\fbeta^{(b)}$ is the estimate from the $b$th subsample. Figures
\ref{fig:1} and \ref{fig:2} present the MSEs of OSP and UNIF, from which we see
the OSP had lower MSEs than UNIF in all scenarios.


\begin{table}[htp] 
  \begin{center}
    \caption{Simulation results of the subsample estimator $\breve{\beta}_1$ with CR = $60\%$.}
    \label{tab:2}
    \vspace{0.1in} \small
    \begin{tabular}{lccccccccccc}
      \hline
      & &  & \multicolumn{4}{c}{OSP} &  & \multicolumn{4}{c}{UNIF} \\
      \cline{4-7}\cline{9-12}
      & $r_k$ & &Bias &ESE &SE &CP & & Bias &ESE &SE &CP\\
      \hline
        Case I
      & 200& & 0.0081 & 0.0839 & 0.0821 &  0.936 &&0.0018 & 0.1075 & 0.1102 & 0.944 \\
      & 400& & 0.0085 & 0.0569 & 0.0571 & 0.946 && 0.0016 & 0.0687 & 0.0769 & 0.978 \\
      & 600& & 0.0048 & 0.0441 & 0.0465 & 0.966 && 0.0005 & 0.0599 & 0.0624 & 0.954 \\
      & 800& & 0.0018 & 0.0393 & 0.0401 & 0.950 &&-0.0015 & 0.0502 & 0.0542 & 0.970 \\
      \hline
      Case II
      & 200& & 0.0048 &  0.0762 & 0.0768 & 0.946 && 0.0027 & 0.1010 &  0.1037 &  0.948 \\
      & 400& & 0.0044 & 0.0526 &  0.0535 & 0.942 && -0.0048 &0.0732 & 0.0734 & 0.942 \\
      & 600& & 0.0042 & 0.0434 & 0.0432 & 0.944 && -0.0041 & 0.0582 & 0.0594 & 0.946 \\
      & 800& & 0.0028 & 0.0352 & 0.0374 &  0.964 && -0.0037 &0.0500 & 0.0518 & 0.962 \\
      \hline
      Case III
      & 200& & 0.0191 & 0.1169 & 0.1216 & 0.954 && 0.0089 & 0.1476 & 0.1569 & 0.962 \\
      & 400& & 0.0179 & 0.0847 & 0.0846 & 0.946 && 0.0030 & 0.1083 & 0.1115 &  0.960 \\
      & 600& & 0.0053 & 0.0691 & 0.0688 & 0.936 && 0.0051 & 0.0889 & 0.0911 &  0.960 \\
      & 800& & 0.0071 & 0.0601 & 0.0592 & 0.930 && 0.0064 & 0.0712 & 0.0794 & 0.960 \\
      \hline
      Case IV
      & 200& & 0.0086 & 0.0509 & 0.0525 & 0.952 && -0.0015& 0.0673 & 0.0829 & 0.982 \\
      & 400& & 0.0019 & 0.0333 & 0.0361 & 0.962 && -0.0016 &0.0466 & 0.0559 & 0.974 \\
      & 600& & 0.0039 & 0.0295 & 0.0292 & 0.944 && -0.0025 &0.0399 & 0.0449 & 0.964 \\
      & 800& & 0.0037 & 0.0243 & 0.0250 & 0.960 && -0.0011 &0.0333 & 0.0380 & 0.978 \\
      \hline
    \end{tabular}
  \end{center}
\end{table}


{\blue The second simulation is conducted to evaluate the computational efficiency of the proposed subsampling method. The same setting is employed to generate random data as in the first simulation, except that $\bbeta_0 = (-1,-0.5,0,0.5,1,0,\cdots,0)^{\prime}$ with $p=5$ and 15. 
The $K$ large survival datasets are denoted as $\mathcal{D}^{[k]} = \{(\X_{ik}, \Delta_{ik}, Y_{ik}), i=1,\cdots,n_k\}$, where the sample size is chosen as $n_k = N/K$. The number of datasets is designated as $K=4$, and the total sample size is set to be $N=10^7$ and $5\times 10^7$, respectively. The calculations were performed using R on a desktop computer equipped with 64GB of memory. We limited the computations to utilize a single CPU core and documented the mean CPU duration based on 10 iterations. In Table~\ref{tab:time},  we report the CPU results for Case I with CR=20\%, where the subsample size is $r_k=800$.  The full data method is also considered for comparison, where $\mathcal{D}_{full}=\cup_{k=1}^K \mathcal{D}^{[k]}$. The computational speed of the OSP estimator is significantly higher than that of the full data estimator using the {\tt coxph} function. The computational burden of the full data method becomes more pronounced with an increase in both the size of the full data sample and the dimensionality of covariates.
The UNIF estimator is computationally faster than the OSP estimator, as it eliminates the need for calculating sampling probabilities. However, it exhibits lower estimation efficiency compared to OSP estimator.}

\begin{table}[H] 
  \begin{center}
    \caption{The CPU time for Case \uppercase \expandafter {\romannumeral 1} with  $r_k=800$ and CR=20\% (in seconds).}
    \label{tab:time}
    \vspace{0.1in} \small
    \begin{tabular}{lllcccccccccc}
      \hline
      & &  & \multicolumn{3}{c}{$p$=5} &  & \multicolumn{3}{c}{$p$=15} \\
      \cline{4-6}\cline{8-10}
      & Sample Size & &UNIF &OSP &Full Data & & UNIF &OSP &Full Data\\
      \hline
   & $N= 10^7$          & &13.28 & 14.78 &102.33  &&19.13 &25.69 &139.70 \\
        & $N=5\times 10^7$   & &72.24 & 79.47 &641.25  &&103.79 &119.32 &913.54  \\
          \hline
    \end{tabular}
  \end{center}
   {\vspace{0cm} \hspace{-0.3cm}\footnotesize $\ddag$ ``Full Data":  calculated with  R function {\tt coxph()}. }
\end{table}

\begin{figure}[H]
  \centering
  \begin{subfigure}{0.45\textwidth}
    \includegraphics[width=\textwidth]{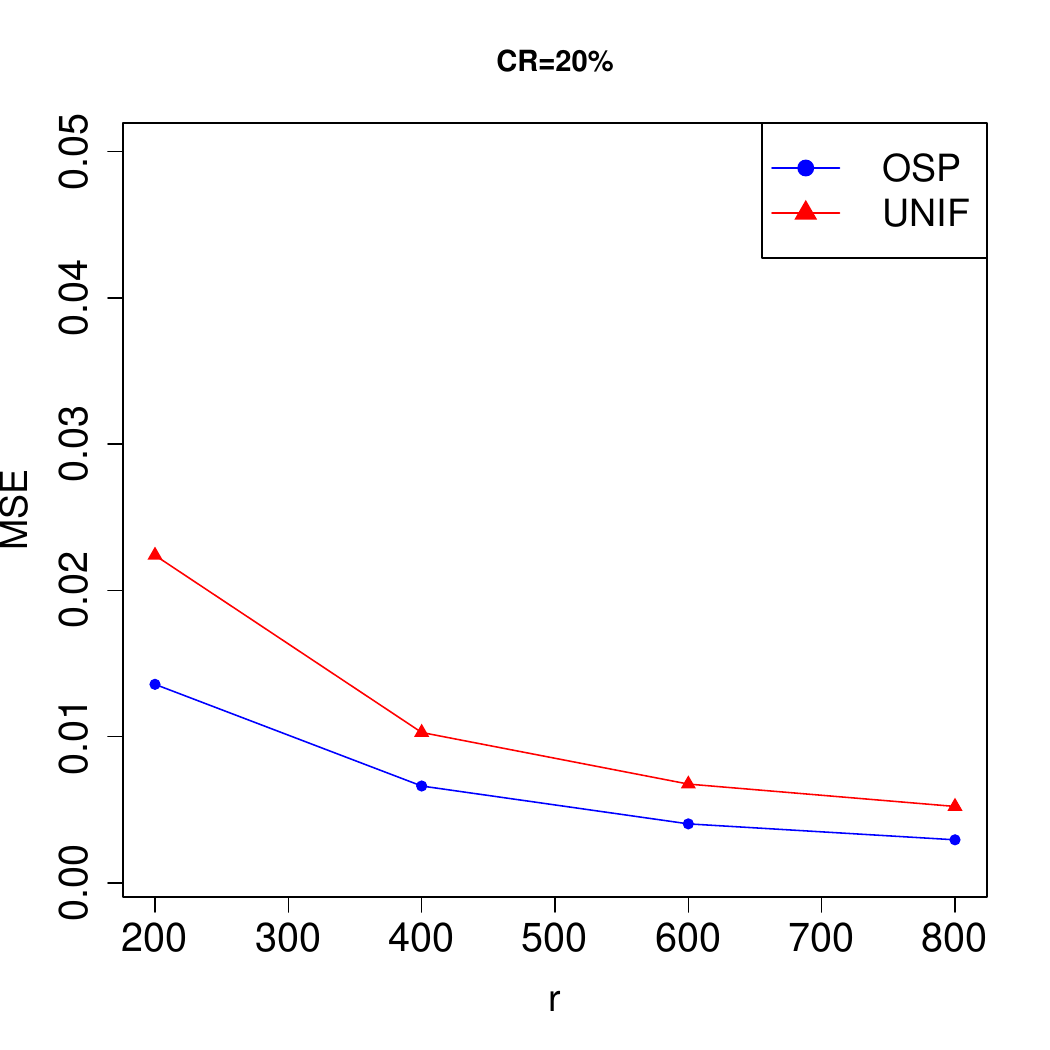}
  \end{subfigure}
  \begin{subfigure}{0.45\textwidth}
    \includegraphics[width=\textwidth]{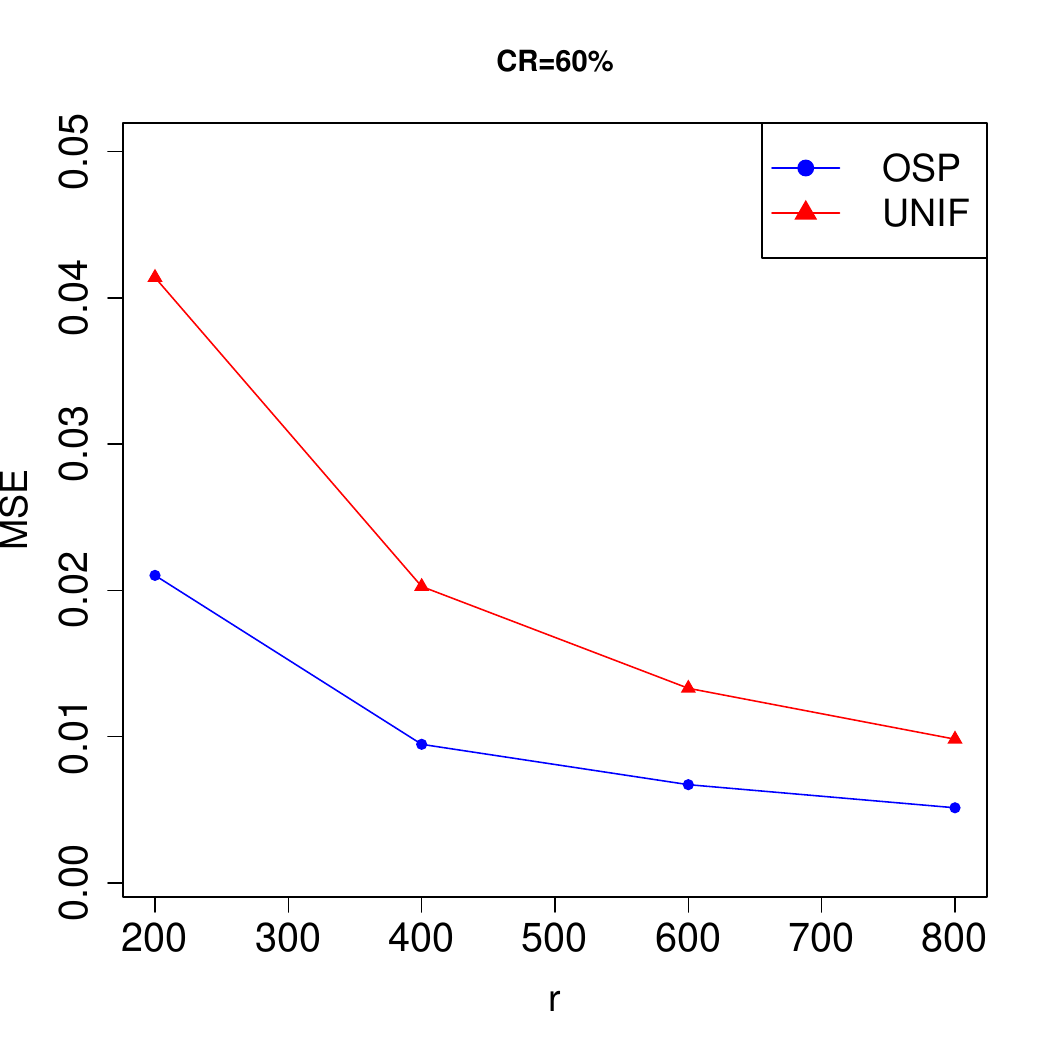}
  \end{subfigure}
 \vspace{-0.1cm}
\begin{center}
\caption{ The MSEs of UNIF and OSP estimators with different covariate distributions across sources.}\label{fig:U-OSP-MX}
\end{center}
\end{figure}

{\blue Lastly, the third simulation is conducted to investigate the performance of subsampling procedure in the presence of heterogeneous covariate distributions across data sources. The random data is generated using the same setup as the first simulation, with the exception that there are different covariate distributions across sources. Specifically, the covariates of $\mathcal{D}^{[1]}$, $\mathcal{D}^{[2]}$, $\mathcal{D}^{[3]}$, and $\mathcal{D}^{[4]}$ are generated from Case I, Case III, Case III, and Case IV respectively. In Table \ref{tab:4-MX}, we 
report the Bias, ESE, SE and CP for $\beta_1$, while other coefficients ($\beta_i$'s) exhibit similar performance and are therefore omitted. The same conclusion can be derived as that of In Tables \ref{tab:1} and \ref{tab:2}, thus aligning with the findings presented in these tables. 
In Figure \ref{fig:U-OSP-MX}, we also present the MSEs of the UNIF and OSP estimators, indicating that the OSP estimator exhibits significantly higher statistical efficiency compared to UNIF. The proposed subsampling method demonstrates satisfactory performance as a whole, even in scenarios where the observed covariates exhibit disparate distributions across sources in our numerical experiment.

}

\begin{table}[H] 
  \begin{center}
    \caption{Simulation results of the subsample estimator $\breve{\beta}_1$ with different covariate distributions across sources.}
    \label{tab:4-MX}
    \vspace{0.1in} \small
    \begin{tabular}{lccccccccccc}
      \hline
      & &  & \multicolumn{4}{c}{OSP} &  & \multicolumn{4}{c}{UNIF} \\
      \cline{4-7}\cline{9-12}
      & $r_k$ & &Bias &ESE &SE &CP & & Bias &ESE &SE &CP\\
      \hline
        CR = $20\%$
      & 200& & 0.0071 & 0.0566 & 0.0571 & 0.946 && -0.0083 & 0.0701 &0.0729 & 0.964 \\
      & 400& & 0.0053 & 0.0388 &0.0393 & 0.950 && -0.0019 &0.0465 & 0.0516 & 0.974 \\
      & 600& &0.0023  & 0.0318 &0.0319 &0.958 && 0.0001 & 0.0391 &  0.0422& 0.954 \\
      & 800& &0.0009  &0.0256 &  0.0276 &0.968 &&-0.0026 &0.0333 &0.0366 & 0.964 \\
      \hline
      CR = $60\%$
      & 200& & 0.0165 & 0.0673 & 0.0713 & 0.956 && -0.0047 &0.0896 & 0.1023 &  0.964 \\
      & 400& & 0.0049 &0.0461 & 0.0493 & 0.956 && -0.0019&  0.0655 & 0.0698 & 0.954 \\
      & 600& & 0.0015 & 0.0392 &  0.0398 & 0.956&&-0.0039 & 0.0547 &  0.0577 & 0.954 \\
      & 800& &0.0043 &0.0347 & 0.0344 & 0.938 && 0.0020 &  0.0463 &  0.0490 & 0.960 \\
      \hline
    \end{tabular}
  \end{center}
\end{table}

\subsection{Real-world Data Example}
In this section, we apply the proposed distributed subsampling
estimation procedure to a large set of USA airline data.  There were 57,729,435
commercial flights that experienced arrival delays within the United States
between October 1987 and April 2008. The failure time is defined as the
duration, measured in minutes,  from the scheduled arrival time ({\blue based on the original departure time}) to the actual
arrival time.   For computational
efficiency, we divided the full samples by decades, and obtained $n_1=
27,221,837$ for data between 1987 and 1997, and $n_2 = 30,507,598$ for data
between 1998 and 2008. In the Cox model, we specified two covariates $X_{ik,1}$
and $X_{ik,2}$ ($k=1, 2$), which represent departure status (0 for on or ahead
of scheduled time and 1 for delayed departure) and distance (a continuous
variable measured in thousands of miles), respectively. It was defined $\X_{ik}
= (X_{ik,1},X_{ik,2})^\prime$. {\blue The full data estimators are also calculated for easy comparison, yielding $\hat{\beta}_1=  -1.0715$ and  $\hat{\beta}_2= -0.1497$.}  {\blue That is to say, the delayed departure status of a flight and its increased airline distance are associated with a decreased ``hazard" of arrival, meaning a prolonged arrival delay time.}


\begin{table}[htp] 
  \begin{center}
    \caption{The Est and (ESE, SE) for subsample estimates in real data.} \label{tab:4}
    \vspace{0.1in}
    \begin{tabular}{llccccccccc}
      \hline
     &  & &OSP &UNIF   \\
      \hline
      $r_k=200$
      &$\beta_1$&   &-1.0690 (0.0983, 0.0937) &-1.0769 (0.1242,  0.1243) \\
      &$\beta_2$&   &-0.1416 (0.0618, 0.0631) & -0.1354 (0.0911, 0.0957) \\
      
      \hline
      $r_k=400$
      &$\beta_1$&  &-1.0654 (0.0638, 0.0657) & -1.0776 (0.0861,  0.0656) \\
      &$\beta_2$&  &-0.1409 (0.0440, 0.0439) & -0.1373 (0.0889,  0.0679) \\
      \hline
      $r_k=600$
      &$\beta_1$&  & -1.0674 (0.0538, 0.0532) & -1.0758 (0.0700,  0.0724) \\
      &$\beta_2$&  &-0.1408 (0.0342,  0.0357) & -0.1427 (0.0529,  0.0549) \\
      \hline
      $r_k=800$
      &$\beta_1$&  &-1.0632 (0.0481, 0.0461) & -1.0652 (0.0597,  0.0633) \\
      &$\beta_2$&  &-0.1402 (0.0304, 0.0308) & -0.1392 (0.0489,  0.0478) \\
      \hline
      $r_k=2500$
      &$\beta_1$&  &-1.0675 (0.0269,  0.0259) & -1.0664 (0.0359,   0.0359) \\
      &$\beta_2$&   &-0.1408 (0.0172, 0.0174) & -0.1398 (0.0262,   0.0271) \\
      \hline
    \end{tabular}
  \end{center}
\end{table}

 For analysis, we consider the OSP
estimator with $\delta=0.1$ and the UNIF estimator, where the pilot subsample
size is chosen as $r_{0k} = 200$.  The values of $r_k$ are set to 200, 400, 600,
800 and 2500.  Table~\ref{tab:4} presents the Est, SE and ESE of subsample-based
estimates based on 500 subsamples, where Est is calculated as the mean of 500 subsampling-based estimators.  Based on the findings presented in
Table~\ref{tab:4}, it can be observed that both estimates derived from
subsampling techniques exhibit consistency as the subsample size
increases. Additionally, the standard error of the OSP estimator is
significantly smaller than that of the UNIF estimator, which confirms the optimality of OSP estimator.

\section{Concluding Remarks}\label{sec6}

In this paper, we proposed a distributed subsampling approach for the
Cox model in massive data settings.  Our approach can alleviate the
computational burden and ensure data privacy during the communication of
multiple datasets. For practical inferences, the asymptotic property of the
proposed estimator was provided.  The simulation study demonstrated that the
proposed approach is effective, and the methodology was applied to analyze a
large dataset from the U.S. airlines.

There are several possible directions for future research. First, we employed
time-independent covariates only but in reality, the covariates can be
time-dependent. It is desirable to extend the proposed approach to incorporate
time-dependent covariates. Second, the proposed framework
lays the foundation for handling other time-consuming tasks with massive
datasets, such as variable selection. Although it has been well known variable
selection with big data is extremely time-consuming, the proposed subsampling
technique offers a promising approach to alleviate the computational
load. Therefore, it would be intriguing to explore the potential of combining the
distributed subsampling estimation approach for Cox models with adaptive
Lasso \cite[]{cox-alasso} and SCAD \cite[]{cox-SCAD}. {\blue Third, distributed learning for the (log) partial likelihood is more subtle than that for linear or generalized linear models since one could not compare the event time of two subjects from two different local sites due to the privacy constraint. In this case, it would be interesting to explore the  fully likelihood approach \cite[]{Full-Cox} for distributed Cox regression via optimal subsampling. Fourth, the extension of our proposed framework to massive survival datasets with rare events is highly desirable. One solution to this issue is the utilization of \cite{rare-Cox-2023}'s subsampling-based estimator and its corresponding Hessian matrix for privacy-preserving data aggregation across local sites. Fifth, the underlying assumption of our method is that the dimensionality of covariates is low, and it requires all observed variables to follow the same distribution across sources. The extension of our framework to scenarios with high-dimensionality and heterogeneity across distributed local data sets remains an unresolved issue, which requires further investigation in our study.}

\section*{Acknowledgements} 
 The authors would like to thank the Editor, the Associate Editor and two reviewers for their constructive and insightful comments  that greatly improved the manuscript.

\section*{Appendix}
\renewcommand{\theequation}{A.\arabic{equation}}
\renewcommand{\thefigure}{A.\arabic{figure}}
\renewcommand{\thetable}{A.\arabic{table}}

\setcounter{equation}{0}
In this section, we give the proof details of Theorems 1 and 2. For presentation clarity, we introduce the following notation:
\begin{align*}
  S_k^{(d)} (t,\bbeta)= \frac{1}{n_k} \sum_{i=1}^{n_k} I(Y_{ik} \geq t) \X_{ik}^{\otimes d}\exp(\bbeta^\prime \X_{ik}), ~d=0, 1~ {\rm or}~ 2,
\end{align*}
where $\mathbf{a}^{\otimes d}$ denotes a power of vector $\mathbf{a}$ with
$\mathbf{a}^{\otimes 0} = 1$, $\mathbf{a}^{\otimes 1} = \mathbf{a}$ and
$\mathbf{a}^{\otimes 2} = \mathbf{a}\mathbf{a}^\prime$. Let $\bar{\mathbf{X}}_k(t,\bbeta) = S_k^{(1)} (t,\bbeta)/S_k^{(0)} (t,\bbeta)$.  Throughout this paper,
$\|\mathbf{A}\| = (\sum_{1\leq i,j \leq p} a_{ij}^2)^{1/2}$ for a matrix
$\mathbf{A}=(a_{ij})$. \\

The derivation of theoretical properties of the distributed subsample estimator $\fbeta$ relies on the following assumptions. 

\begin{assumption}\label{assu1}
  The baseline hazard satisfies that $\int_0^\tau \lambda_0(t)dt < \infty$, and
  $P(T_{ik} \geq \tau)>0$. 
\end{assumption}
\begin{assumption}\label{assu2}
  For $k=1,\cdots,K$, the quantity
  $\frac{1}{n_k}\sum_{i=1}^{n_k}\int_0^\tau\Big[\frac{S_k^{(2)}(t,{\bbeta})}{S_k^{(0)}(t,
    \bbeta)} - \left\{\frac{S_k^{(1)}(t, \bbeta)}{S_k^{(0)}(t,
      \bbeta)}\right\}^{\otimes 2} \Big]dN_{ik}(t)$ converges in probability to
  a positive definite matrix for all $\bbeta\in\Theta$, where $\Theta$ is a
  compact set containing the true value of $\bbeta$.
\end{assumption}
\begin{assumption}\label{assu3}
  The covariates $\mathbf{X}_{ik}$'s are bounded.
\end{assumption}
\begin{assumption}\label{assu5}
  For $k=1,\cdots,K$, there exists two positive definite matrices,
  $\mathbf{\Lambda}_1$ and $\mathbf{\Lambda}_2$ such that
  \begin{eqnarray*}
    \mathbf{\Lambda}_1 \leq \frac{1}{n_k}\sum_{i=1}^{n_k}\int_0^\tau\left[\frac{S_k^{(2)}(t,{\bbeta_0})}{S_k^{(0)}(t,
    \bbeta_0)} - \left\{\frac{S_k^{(1)}(t, \bbeta_0)}{S_k^{(0)}(t,
    \bbeta_0)}\right\}^{\otimes 2} \right]dN_{ik}(t) \leq \mathbf{\Lambda}_2.
  \end{eqnarray*}
  i.e., for any $\mathbf{v}\in \mathbb{R}^p$,
  $\mathbf{v}^\prime\mathbf{\Lambda}_1\mathbf{v} \leq
  \frac{1}{n_k}\sum_{i=1}^{n_k}\int_0^\tau\mathbf{v}^\prime\Big[\frac{S_k^{(2)}(t,{\bbeta_0})}{S_k^{(0)}(t,
    \bbeta_0)} - \left\{\frac{S_k^{(1)}(t, \bbeta_0)}{S_k^{(0)}(t,
      \bbeta_0)}\right\}^{\otimes 2} \Big]\mathbf{v}dN_{ik}(t) \leq
  \mathbf{v}^\prime\mathbf{\Lambda}_2\mathbf{v}$.
\end{assumption}

{\blue Assumptions \ref{assu1} through \ref{assu3} are three commonly imposed conditions
on the Cox's model {(\citeauthor{Andersen1982}, \citeyear{Andersen1982};
  \citeauthor{Huang-J2013-AOS}, \citeyear{Huang-J2013-AOS};
  \citeauthor{Fang2016-JRSSB}, \citeyear{Fang2016-JRSSB}).}  Assumption \ref{assu5} is required to establish the
asymptotic normality of distributed subsample estimator \cite[]{Lin-DC-2011}.}

 For the purpose of enhancing clarity in presenting our subsampling procedure, we provide the asymptotic properties of subsample-based estimators $\tbeta_k$'s, which can be derived analogously to Proposition 1 in \cite{JCGS-Cox}, along with Proposition 2 from \cite{IEEE-Poisson}.

\begin{lemma}\label{Lemma1}
  Under the assumptions~\ref{assu1}-\ref{assu3}, $r_k = o(n_k)$, as $n_k\rightarrow \infty$ and
  $r_k\rightarrow \infty$, then the $k$th subsample-based estimator $\tbeta_k$
  is a consistent estimator of $\bbeta_0$ with a convergence rate
  $O_{P}(r_k^{-1/2})$, where $k=1,\cdots,K$.  In addition, we
  have
  \begin{align}\label{Eq_sigmak}
    \bSigma_k^{-1/2}(\tbeta_k - \bbeta_0) \stackrel{d}{\longrightarrow} N(\mathbf{0},\mathbf{I}),
  \end{align}
  where $\stackrel{d}{\longrightarrow}$ denotes convergence in distribution, 
  $\bSigma_k =\mathbf{\Psi}_k^{-1}\mathbf{\Gamma}_k\mathbf{\Psi}_k^{-1}$,
  \begin{align}\label{Eq5}
    \mathbf{\Psi}_k = \frac{1}{n_k}\sum_{i=1}^{n_k}\int_0^\tau\left[\frac{S_k^{(2)}(t,\bbeta_0)}{S_k^{(0)}(t, \bbeta_0)} -  \left\{\frac{S_k^{(1)}(t, \bbeta_0)}{S_k^{(0)}(t, \bbeta_0)}\right\}^{\otimes 2}  \right]dN_{ik}(t) 
  \end{align}
with $N_{ik}(t) = I(\Delta_{ik} =1, Y_{ik} \leq t)$,  and
  \begin{eqnarray}\label{Eq6}
    \mathbf{\Gamma}_k &=&\frac{1}{n_k^2} \sum_{i=1}^{n_k} \frac{1}{\pi_{ik}} \left[\int_0^\tau\left\{\X_{ik} - \bar{\X}_k(t, \bbeta_0)\right\}dM_{ik}(t,\bbeta_0)\right]^{\otimes 2}\nonumber\\
                      &&- \frac{1}{n_k^2}\sumn \left[\int_0^\tau\left\{\X_{ik} - \bar{\X}_k(t, \hbeta)\right\}dM_{ik}(t,\hbeta)\right]^{\otimes 2}
  \end{eqnarray}
  with $M_{ik}(t,\bbeta) = N_{ik}(t) - \int_0^t I(Y_{ik} \geq u)\exp(\bbeta^\prime
\X_{ik})\lambda_0(u)du$.
\end{lemma}

\vspace{0.5cm}

{\noindent{{\bf Proof of Theorem~\ref{Th4}}}}.  Under the assumptions~\ref{assu1}-\ref{assu5} and $r_k = o(n_k)$, the asymptotic normality presented in  (\ref{Eq_sigmak}) indicates that the variable $\breve{\bbeta}_k$ asymptotically follows a independent normal vector with mean $\hbeta$ and covariance matrix $\bSigma_k$, where $k=1,\cdots,K$. Based on Lemma \ref{Lemma1}, we have the following expression:
\begin{eqnarray}\label{S-A79}
\bPsi_k(\breve{\bbeta}_k - \hbeta)= \mathbf{Z}_k + \bPsi_kR_k, 
\end{eqnarray}
where $\mathbf{Z}_k$ represents a normal random vector with mean zero and covariance matrix $\mathbf{\Gamma}_k$, and $R_k=O_P(r_k^{-1})$.  By taking summation over $k$ on both side of
Equation (\ref{S-A79}), we get
\begin{eqnarray}\label{AS7-80}
\left(\sum_{k=1}^K\bPsi_k\right)^{-1}\sum_{k=1}^K \bPsi_k\breve{\bbeta}_k - \hbeta = \left(\sum_{k=1}^K\bPsi_k\right)^{-1} \sum_{k=1}^K \mathbf{Z}_k + \left(\sum_{k=1}^K\bPsi_k\right)^{-1}\sum_{k=1}^K\bPsi_kR_k. 
\end{eqnarray}
Let $\lambda_1 > 0$  be the smallest eigenvalue of the matrix $\mathbf{\Lambda}_1$, and ${\lambda}_2$ be the
largest eigenvalue of the matrix  $\mathbf{\Lambda}_2$, where $\mathbf{\Lambda}_1$ and $\mathbf{\Lambda}_2$ are given in the assumption \ref{assu5}. Then for any vector $\mathbf{v}\in \mathbb{R}^p$, we have $\mathbf{v}^\prime\bPsi_k\mathbf{v}\geq \mathbf{v}^\prime\mathbf{\Lambda}_1\mathbf{v}\geq\lambda_1\|\mathbf{v}\|^2$, which is due to the assumption \ref{assu5}. Hence, $\mathbf{v}^\prime\frac{1}{K}\sum_{k=1}^K\bPsi_k\mathbf{v}\geq \lambda_1\|\mathbf{v}\|$. In the remainder of proof, the norm of a $p\times p$ positive definite matrix $\mathbf{A}$ is defined as $\|\mathbf{A}\|=\sup_{\mathbf{v}\in \mathbb{R}^p}\frac{\|\mathbf{A}\mathbf{v}\|}{\|\mathbf{v}\|}$. Based on the Facts A.1 and A.2 of \cite{Lin-DC-2011}, we have
\begin{eqnarray*}
\left\|\left(\frac{1}{K}\sum_{k=1}^K\bPsi_k\right)^{-1}\right\|\leq \frac{1}{\lambda_1}.
\end{eqnarray*}
This together with $\|\bPsi_k\|\leq \|\mathbf{\Lambda}_2\| \leq \lambda_2$ lead to that
\begin{eqnarray}
\left\|\left(\sum_{k=1}^K\bPsi_k\right)^{-1}\bPsi_k\right\|\leq \left\|\left(\frac{1}{K}\sum_{k=1}^K\bPsi_k\right)^{-1}\right\|\left\|\frac{1}{K}\bPsi_k\right\|
\leq \frac{\lambda_2}{K\lambda_1}.
\end{eqnarray}
Hence, as $r_k \rightarrow \infty$ the last term of (\ref{AS7-80}) satisfies
\begin{eqnarray*}
\left\|\left(\sum_{k=1}^K\bPsi_k\right)^{-1}\sum_{k=1}^K\bPsi_kR_k\right\|&\leq&
\sum_{k=1}^K\left\|\left(\sum_{k=1}^K\bPsi_k\right)^{-1}\bPsi_kR_k\right\|\\
&\leq& \frac{\lambda_2}{\lambda_1 K}\sum_{k=1}^K \|R_k\|\\
& = & o_P(1).
\end{eqnarray*}
 Therefore, as $r_k \rightarrow \infty$ we have 
\begin{align}\label{A.80}
\left(\sum_{k=1}^K\bPsi_k\right)^{-1}\sum_{k=1}^K \bPsi_k\breve{\bbeta}_k - \hbeta \stackrel{d}{\longrightarrow} N\left(0,\left\{\sum_{k=1}^K\bPsi_k\right\}^{-1}\left\{\sum_{k=1}^K{\boldsymbol\Gamma}_k\right\}\left\{\sum_{k=1}^K\bPsi_k\right\}^{-1}\right).
\end{align}
As $\breve{\mathbf{\Psi}}_k$ and $\breve{\mathbf{\Gamma}}_k$ are consistent to ${\mathbf{\Psi}}_k$ and ${\mathbf{\Gamma}}_k$, respectively,   the application of Slutsky's theorem in conjunction with  (\ref{A.80}) guarantees that
\begin{align*}
    \breve{\boldsymbol \Omega}_{\mbox{\tiny \rm  DSE}}^{-1/2}(\fbeta - \bbeta_0) \stackrel{d}{\longrightarrow} N(\mathbf{0},\mathbf{I}),
\end{align*}
  where $\breve{\boldsymbol \Omega}_{\mbox{\tiny \rm  DSE}} = (\sum_{k=1}^K \breve{\mathbf{\Psi}}_k)^{-1} (\sum_{k=1}^K \breve{\mathbf{\Gamma}}_k)(\sum_{k=1}^K \breve{\mathbf{\Psi}}_k)^{-1}$. This ends the proof.

\vspace{1cm}
\noindent{\bf Proof of Theorem \ref{Theorem2}}.    Subtracting $\bbeta_0$ from both sides of (\ref{fed-est}), we can derive that
\begin{eqnarray*}
\fbeta - \bbeta_0 = \left(\sum_{k=1}^K \breve{\mathbf{\Psi}}_k\right)^{-1}\left[\sum_{k=1}^K \breve{\mathbf{\Psi}}_k(\breve{\bbeta}_k-\bbeta_0)\right].
\end{eqnarray*}
Therefore,
\begin{eqnarray*}
\|\fbeta - \bbeta_0\| &\leq& \sum_{k=1}^K \left\|\left(\sum_{k=1}^K \breve{\mathbf{\Psi}}_k\right)^{-1}\breve{\mathbf{\Psi}}_k(\breve{\bbeta}_k-\bbeta_0)\right\|\\
 &\leq& \sum_{k=1}^K \|\breve{\bbeta}_k-\bbeta_0\|,
\end{eqnarray*}
where the last inequality is due to $\|(\sum_{k=1}^K \breve{\mathbf{\Psi}}_k)^{-1}\breve{\mathbf{\Psi}}_k\|\leq 1$. Let $k_0 = \arg\max_{1\leq k \leq K} \{\|\breve{\bbeta}_{k} - \bbeta_0\|\}$, then we have 
 \begin{eqnarray*}
\|\fbeta - \bbeta_0\|\leq K \|\breve{\bbeta}_{k_0} - \bbeta_0\|.
 \end{eqnarray*}
This completes the proof.

\bibliographystyle{natbib}
\bibliography{reference}
\end{document}